\documentstyle[aps,pra,epsbox]{revtex}
\begin{document}
\draft
%\preprint{kasei-hep/000931}
\title{Gauge Symmetry Breakdown \\
due to Dyanamical versus Elementary Higgs}

\author{Takayuki Matsuki
\thanks{E-mail: matsuki@tokyo-kasei.ac.jp}}
\address{Tokyo Kasei University,
1-18-1 Kaga, Itabashi, Tokyo 173-8602, JAPAN}
\author{Masashi Shiotani
\thanks{E-mail: shiotani@tanashi.kek.jp}}
\address{Faculty of Science and Technology,
Kobe University, 1-1 Rokkohdai,\\ Nada, Kobe 657, JAPAN}
\maketitle

\begin{abstract}
We study in details on how gauge bosons can acquire mass when the chiral
symmetry dynamically breaks down for massless gauge theory without scalars.
Introducing dynamical scalar fields into the original gauge theory, we show
that when the chiral symmetry breaks down, the theory gives gauge boson
masses different from what would be obatained if an elemetary Higgs is
included. We clarify the reason and propose one method how to calculate gauge
boson masses in the case of dynamical gauge symmtry breakdown. We explain the
method by using an example in which $SU(5)$ massless gauge theory breaks down
to $SU(4)$ with massless fermions in appropriate representations.

\end{abstract}
\medskip
%%%%%%%%%%%%%%%%%%%%%%%%%%%%%%%%%%%%%%%%%%%%%%%%%%%%%%%%%%%%
%    Introduction
%%%%%%%%%%%%%%%%%%%%%%%%%%%%%%%%%%%%%%%%%%%%%%%%%%%%%%%%%%%%
\section{Introduction}
\label{intro}
The method that people normally use to calculate gauge boson masses from
massless gauge theory without an elementary scalar is to apply the
Jackiw-Johnson sum rule\cite{JJ}, which gives an expression, $g\; f$, i.e., a
gauge coupling constant times a decay constant. The latter quantity, $f$, is
more explicitly given by the Pagels-Stokar formula.\cite{PS}

When this method is combined with the idea of the tumbling proposed by Raby,
Dimopoulos, and Susskind\cite{RDS,MP}, people had thought they have everything
in their hands, namely how to obtain the breaking orientation of the gauge
group, how to obtain fermion masses, and how to obtain gauge boson masses.
Actually the developements of dynamical theories since then, like theories of 
technicolor, extended-technicolor, etc., solely assume this expression,
$M_G=g\,f$. Also it is taken for granted that the results of technicolor
theory\cite{WS} can be in principle derived by a theory replacing the
condensation with an elementary Higgs scalar even though technicolor theory
gives some restriction on model building. {\sl Our question we would like to
ask in this paper is whether gauge boson masses should be calculated after the
breaking direction is determined by fermion condensation. Or is it possible
to determine both fermion and gauge boson masses at the same time once the
condensation pattern is determined? Is there really no difference between
theories with and without scalars?}

Let us recall what the idea of tumbling gives us when constructing a dynamical
model. Assumption of the tumbling is supposed to give us the scenario how
gauge symmetry dynamically breaks down in series, once an original massless
gauge theory with massless fermions but Higgs scalars is given. The
direction of breakdown is selected by the so-called most attractive channel
(MAC), which can be determined from the second Casimir invariants. Due to the
chiral symmetry breakdown, fermions in a certain representation acquire mass
and tacitly assumed is that at the same time some gauge bosons also acquire
mass since the gauge symmetry breaks down when the chiral symmetry is broken by
a gauge-non-singlet dynamical scalar which is a composite of
fermion-anti-fermion pair. What we would like to study in this paper is on this
point. That is, when the chiral symmetry spontaneously breaks down, we shall
calculate the decay constants of the Nambu-Goldstone (NG) bosons which are
abosorbed by would-be massive gauge bosons. These decay constants are nothing
but gauge boson masses up to coupling constants. After this calculation we will
compare the results with those expected from a theory with an elementary Higgs
scalar and see whether they are equivalent or not.\cite{REV}

In Sec. \ref{tumbling}, by adopting the massless $SU(5)$ gauge theory as an
example, we will show what is the problem and where is the problem when one
wants to calculate gauge boson masses. In the same section, we will also point
out that at least in our model calculation, theories with and without elementary
Higgs scalars are not equivalent to each other. In Sec. \ref{su5}, we will
propose one solution how to calculate gauge boson masses and apply it to our
case. In Sec. \ref{mass}, mass formulae both for fermions and gauge bosons are
given for the massless $SU(5)$ gauge theory without scalar and we find that one
arbitrary parameter comes in for gauge boson masses. That is, one of gauge boson
masses cannot be determined uniquely {\it a la} Pagels and Stokar. In Sec.
\ref{discuss}, the results are compared with those of a theory with an
elementary Higgs and discussions on the results are also given. Detailed
calculations are given in the Appendix.

%%%%%%%%%%%%%%%%%%%%%%%%%%%%%%%%%%%%%%%%%%%%%%%%%%%%%%%%%%%%
%    Tumbling Scenario
%%%%%%%%%%%%%%%%%%%%%%%%%%%%%%%%%%%%%%%%%%%%%%%%%%%%%%%%%%%%
\section{Tumbling Scenario}
\label{tumbling}

We start from the massless $SU(5)$ gauge theory whose tree action without
scalar fields is given by
\begin{equation}
  {\cal S}_0 = \int d^4x\,\Big\{
  i\left(\overline\psi\right)^i\left[{\partial\kern-0.6em /}\;\delta_i{}^j-
  ig{A^A\kern-1.2em /}{\kern+0.7em}\left(T^A\right)_i{}^j\right]\psi_j
  +\frac{i}{2}{\left(\overline\chi\right)}{~}^{i\;j}\left[
  {\partial\kern-0.6em /}{\kern+0.4em}\delta_j{}^k-2ig{A^A\kern-1.2em /}
  {\kern+0.5em}\left(T^A\right)_j{}^k\right]\chi_{k\;i}
  -\frac{1}{4}\left(F^A_{\mu\;\nu}\right)^2 \Big\},\label{S0}
\end{equation}
where an index $A$ for gauge fields and generators runs from 1 through 24 and
representations of particle multiplets under $SU(5)$ are given by
\begin{mathletters}
\begin{eqnarray}
  \left({\psi^c}\right)^i_L &&\quad{\rm :~~~~}\underline{5}^*, \\
  \left(\chi_{i\;j}\right)_L &&\quad{\rm :~~~~}\underline{10}, \\
  \left(A_\mu\right)_i{~}^j=\sum_{A=1}^{24}A^A_\mu \left(T^A\right)_i{~}^j &&
  \quad{\rm :~~~~}\underline{24}.
\end{eqnarray}
\end{mathletters}
Fermion kinetic terms in Eq.~(\ref{S0}) include $\psi _i\equiv(\psi _i)_R$
which corresponds to $\underline{5}$, while fermions with bar on the top
behave like their conjugates, e.g., $\left(\overline\psi\right)^i\sim
\underline{5}^*$ and $\left(\overline\chi\right)^{i\;j}\sim\underline{10}^*$.
Note also that
\[
  \left(\overline\chi\right)^{i\;j}=\overline{\chi{}^{~}_{i\;j}}.
\]
First we integrate over all the gauge fields in Eq.~(\ref{S0}) to obtain
effective four-fermi interactions ($S_1$),
\begin{eqnarray}
  S_1&=&\int d^4x \left[i\,\left(\overline\psi\right)^i
  {\partial\kern-0.6em /}\;\delta{}_i{}^j \psi_j
  +i\;\frac{1}{2}\left(\overline \chi\right)^{i\;j}
  {\partial\kern-0.6em /}{\kern+0.4em}\delta_j{}^k \chi_{k\;i}\right]
%  \nonumber \\
  + \frac{1}{2}g^2\int d^4x\;d^4y\;\Big[{J_5}^A_\mu(x)
  D^{\mu\;\nu}(x-y){J_5}^A_\nu(y)
  \nonumber \\
  &+&{J_{10}}^A_\mu(x) D^{\mu\;\nu}(x-y){J_{10}}^A_\nu(y)
  +2{J_5}^A_\mu(x) D^{\mu\;\nu}(x-y){J_{10}}^A_\nu(y)\Big]
  +\ldots, \label{S1}
\end{eqnarray}
where
\begin{mathletters}
\begin{eqnarray}
  {J_5}^A_\mu (x)&=&{\overline\psi}{~}^i(x)\gamma_\mu \left(T^A\right)_i{}^j
  \;\psi_j(x), \\
  {J_{10}}^A_\mu (x)&=&{\left(\overline\chi\right)}{~}^{i\,k}(x)\;\gamma_\mu 
  \left(T^A\right)_k{}^\ell\;\delta{}_i{}^j\,\chi{}_{\ell\,j}(x), \\
  D_{\mu\nu}(x)&=&\int\frac{d^4p}{(2\pi)^4}\frac{1}{p^2}
  \left[g_{\mu\nu} -(1-\alpha_0)\frac{p_\mu p_\nu}{p^2}\right]\;e^{-ipx},
\end{eqnarray}
\end{mathletters}
with $i$, $j$, $k$, and $\ell$ running from 1 through 5, and $\alpha_0$ being a
gauge parameter for $SU(5)$ in the last equation.

The MAC (there recently occurs some objection to the MAC arguments.\cite{HR})
criterion which selects most preferable condensation corresponds to
which coefficient is the largest when Fierz transformation of these
four-fermi interactions is performed. We then find that the coefficient of
scalar component of the Fierz-transformed $J_{10}\times J_{10}$ is the largest
among all the current products and obtain an effective action $S_2$ with
non-local bilinear interaction terms,
\begin{eqnarray}
  &{~}&{J_{10}}^A_\mu (x){J_{10}}^A_\nu (y)
  \nonumber \\
  &=&-\frac{1}{8}\;{\rm Tr}\;\left[{\left(\overline\chi\right)}{~}^{i\;k}(x)
  \gamma_\mu \left(T^A\right)_k{}^\ell\;\delta{}_i{}^j\,\chi_{\ell\;j}(x)\right]
  \;{\rm Tr}\;\left[\left({\tilde\chi{}^{m\;n\;o}(y)}^T\;
  U_c\right)\;\gamma_\nu\left(T^A\right)_o{}^p\;\delta{}_n{}^q\,
  \left(U_c^{-1}{\overline
  {\tilde\chi}}{}_{pqm}(y)^T\right)\right]
  \nonumber \\
  &=& \frac{2C_{10}-C_5}{64N_5} \eta_{\mu\nu}\;
  {\rm Tr}\;\left[\left({{\tilde\chi}{~}^{i\,j\,p}}(y)^T\;U_c\right)
  \chi_{p\,j}(x)\right]\;
  {\rm Tr}\;\left[{\left(\overline\chi\right)}{~}^{k\,q}(x)\;\left(U_c^{-1}\;
  {\overline{\tilde\chi}}{}_{q\,k\,i}(y)^T\right)\right]
  +\ldots, \label{J10J10}
\end{eqnarray}
where the Fierz transformation for ten dimensional generators, 
${\left(T^A\right)_i{}^j}_k{}^\ell$, is used and defined are
\begin{equation}
  \tilde\chi{}^{ijk}(x)=\epsilon^{ijk\ell m}\chi_{\ell m}(x), \qquad
  U_c=i\gamma^0\gamma^2.
\end{equation}
The coefficient of Eq.~(\ref{J10J10}) is the same result as the tumbling
scenario, which was also derived by using an effective potential approach
\cite{CJT} by Kikukawa and Kitazawa.\cite{KK} They have also considered
contributions of massive gauge bosons to the fermion gap equation in the
next sequence of the tumbling with regard to the gauged Nambu-Jona-Lasinio
model. Namely they have neither pondered over nor calculated gauge boson
masses at the {\sl first time the chiral symmetry is spontaneously broken.}
We would like to go further on to show where the obstacle is when calculating
gauge boson masses at the first time the chiral symmetry is broken and to show
that Eq.~(\ref{S1}) with Eq.~(\ref{J10J10}) is not the one we start from.

Introducing a Gaussian term of auxiliary fields so that it cancels four-fermi
terms given by Eq.~(\ref{J10J10}), we finally obtain an effective action $S_3$
with Yukawa interaction terms as follows.
\begin{eqnarray}
  S_3&=&S_2+S_{\rm AF}
  \nonumber \\
  &=&\int d^4x \left[i\;\left(\overline\psi\right)^i
  {\partial\kern-0.6em /}\;\delta_i{}^j \psi_j
  +i\;\frac{1}{2}\left(\overline \chi\right)^{i\;j}
  {\partial\kern-0.6em /}{\kern+0.4em}\delta_j{}^k \chi_{k\;i}\right]+
  \frac{1}{2}\int\,d^4x\,d^4y\,\Bigl[\left({{\tilde\chi}{~}^{i\,\ell\,j}}
  (y)^T\;U_c\right) \phi_i(x,y)\;\delta{}_j{}^k \chi_{k\,\ell}(x)
  + {\rm H.C.}\Bigr]
  \nonumber \\
  &-& \frac{1}{2}\int\,d^4x\,d^4y\,
  {\phi_i}{}^\dagger(x,y)\;D_0^{-1}(x-y)\;\phi_i(x,y)
  + \ldots, \label{S3}
\end{eqnarray}
where only relevant terms are written, and defined are
\begin{mathletters}
\begin{eqnarray}
  D_0(x-y)&=&\lambda_0\,D(x-y), \qquad
  \lambda_0=\frac{1}{32N_5}\left(2C_{10}-C_5\right), \\
  D(x-y)&=&\frac{g^2}{4}\,g^{\mu\,\nu}D_{\mu\,\nu}(x-y)=
  \frac{3+\alpha_0}{4}\,g^2\int\frac{d^4p}{\left(2\pi\right)^4}
  \frac{e^{-ip(x-y)}}{p^2}=-\frac{(3+\alpha_0)g^2}{16\pi^2}
  \frac{1}{\left(x-y\right)^2},
\end{eqnarray}
\end{mathletters}
with $\alpha_0$ being a gauge parameter and the function $D^{-1}_0(x-y)$ being
equal to $1/D_0(x-y)$. The constants $C_i$ are the second rank Casimir
invariants for $i$ dimentions and are given by $C_{10}=18/5$ and $C_5=12/5$.
Here as a summary we have derived that the MAC is the ${\underline 5}\in
{\underline 10}\times {\underline 10}$ channel,
$\phi_i\sim \epsilon^{ijk\ell m}\chi_{jk} \chi_{\ell m}$, by demanding that a
coupling be the largest after Fierz-transforming all the current products,
which is the same results as those of Ref. \cite{RDS}, of course, so that only
a composite scalar $\phi_i$ among many auxiliary fields is written in
Eq.~(\ref{S3}).

In Eq.~(\ref{S3}) we may generally assume that $\phi_i(x,y)$ acquires a VEV as
\begin{eqnarray}
  \phi_i(x,y) &=& \left\{\frac{\phi_0(x-y)}{v}\left[v+\frac{1}{\sqrt{2}}\;
  \sigma\left(\frac{x+y}{2}\right)\right]\exp\left[\frac{i\pi^\alpha
  ((x+y)/2) T^\alpha }{v}\right]\eta\right\}_i
  \nonumber \\
  &=& \frac{\phi_0(x-y)}{v}\left[v+\frac{1}{\sqrt{2}}\;
  \sigma\left(\frac{x+y}{2}\right)\right]\delta_i{}^5
  +i\frac{\phi_0(x-y)}{v}\pi^\alpha \left(\frac{x+y}{2}\right)
  \left(T^\alpha \right)_i{}^5+\ldots, \label{phi0} \\
  \eta_i &\equiv& \delta_i{}^5, 
  \nonumber
\end{eqnarray}
where assumed is a decomposition of a bilocal field as a product of local
fields\cite{MS}. The translationally invariant vacuum expectation value (VEV)
$\phi_0(x-y)$ of $\phi(x,y)$ becomes a running fermion mass when expressed in a
momentum space. Then legitimacy of decomposition given by Eq.~(\ref{phi0}) is
assured by checking masslessness of the NG bosons, $\pi^\alpha$.\cite{MS} The
VEV given by Eq.~(\ref{phi0}) breaks the original $SU(5)$ symmetry to $SU(4)$.
Though we have functionally integrated out all the $SU(5)$ gauge fields, we
still can consider $SU(5)$ gauge fields attached to the external
legs in the Feynman diagrams since Eq.~(\ref{S3}) is obtained by taking only
internal gauge boson loops into account. That how many NG bosons should be
introduced in Eq.~(\ref{phi0}) depends on how the gauge symmetry breaks down.
We now know that $SU(5)$ breaks down to $SU(4)$, hence nine NG bosons are
necessary corresponding to the 9 ($=1+4+4^*$) would-be massive gauge bosons.
Hence superscript $\alpha$ of $\pi^\alpha$ and $T^\alpha$ in Eq.~(\ref{S3})
runs from 16 through 24.

\begin{center}
%
% Figure 1
%
\begin{figure}
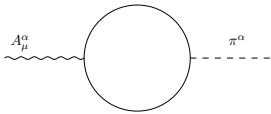

\center\psbox[hscale=0.5,vscale=0.5]{nfig1.eps}
\caption{Fermion one-loop contribution to $A_\mu^\alpha$ - $\pi^\alpha$
two point function.}
\label{Fig1}
\end{figure}
\end{center}

Now we can calculate mass for the gauge fields which absorb the NG bosons
generated due to the chiral symmetry breakdown. The chiral symmetry breakdown
induces gauge symmetry breakdown since $\phi_i$ which breaks chiral symmetry is
5-dimensional. The gauge symmetry breakdown is induced by the fermion one-loop
diagrams shown in Fig.~\ref{Fig1} since there are no elelmentary couplings
between gauge fields and NG bosons. Gauge boson mass is obtained by calculating
these diagrams.

These are the procedures expected by Raby, Dimopoulos, and Suskind and other
followers, for instance see Ref. \cite{KK}. Namely people expected the would-be
massive gauge bosons have masses with the same ratio as expected from an
elementary Higgs. However let us carefully analyze this diagram. When the 10
dimensional fermion multiplet breaks down to $6+4$ under $SU(4)$, the fermion
loop consists of 6 and 6 if the singlet gauge field and NG boson are
externals but the loop consists of $4/4^*$ and 6 if the four-dimensional gauge
field and NG boson are externals. That is, the loop structures are different if
the external gauge fields/NG bosons are in different representations. Hence
accordingly the decay constants defined by Fig.~\ref{Fig1} have different
expressions for singlet and four-dimensional gauge fields. That is, even though
we have assumed symmetry breakdown takes place due to the dynamical Higgs given
by Eq.~(\ref{phi0}), it is not a correct expression at all. The VEV's or decay
constants involved with singlet and four-dimentsional NG bosons must be
different functionals of $\phi_0(x-y)$ in our case, which means the ratio
between singlet and quartet gauge boson masses is not the same constant value as
expected from an elementary Higgs.

Another complexity comes in when one considers gauge boson mass together with
fermion mass. When one considers the gap equation for 6-dimensional fermion
multiplet under $SU(4)$, all the $SU(5)$ gauge fields are taken into account,
massless as well as massive. In this case this gap equation assures masslessness
of all the nine NG bosons $\pi^\alpha$ in Eq.~(\ref{phi0}) and it has the same
coefficient as in Ref.~\cite{KK} expected from the MAC, however, normally the
gap equation for fermion only takes into account massless gauge fields if we
neglect the gauged Nambu-Jona-Lasinio consideration in \cite{KK}. So somehow we
have to get rid of massive gauge fields from the gap equation since their
contribution is as the order of 1/(gauge boson mass)$^2$. Even if we succeed in
excluding contributions from massive gauge fields, the fermion loop structures
of decay constants are different between singlet and four-dimensional would-be
massive gauge fields as above.

We can now answer to the questions raised in Sect. \ref{intro}. {\sl Gauge boson
masses should be calculated at the same time when the gap equation for the
fermions takes only massless gauge fields into account. There is difference
between theories with dynamical and elementary Higgs scalars.}

The VEV of a composite scalar $\phi_i(x,y)$, $\phi_0(x-y)$, is the only mass
dimensionful qunatity appearing in the theory. The VEV of a local scalar has a
constant value, $v$, given in Eq.~(\ref{phi0}) which is defined as a decay
constant and a functional of $\phi_0(x-y)$ as shown by Pagels and Stokar
\cite{PS}. However, there are two kinds of decay constants in our case and
these are different functionals of $\phi_0(x-y)$. Since there are no elementary
couplings between gauge fields and NG bosons, fermion one-loop diagrams that
generate these couplings have different stuructures. We will see in
Sec.~\ref{mass} that not all the decay constans can be calculated {\it a la}
Pagels and Stokar. We need another principle to calculate the decay constant.
In the next Section we will show one method how to treat appropriately this
situation.

%%%%%%%%%%%%%%%%%%%%%%%%%%%%%%%%%%%%%%%%%%%%%%%%%%%%%%%%%%%%
%    Effective Action for SU(5)
%%%%%%%%%%%%%%%%%%%%%%%%%%%%%%%%%%%%%%%%%%%%%%%%%%%%%%%%%%%%
\section{Effective Action}
\label{su5}
Based on the former observations, in this section we will explain how to obtain
the effective action which generates appropriate Feynman rules for elementary
fermions, massive gauge fields, and dynamical Higgs fields so that contributions
of massive gauge fields are excluded from the fermion gap equation by using an
example of $SU(5)$ massless gauge theory and assuming gauge symmetry breakdown
$SU(5)\rightarrow SU(4)$ which is the result of the MAC hypothesis. 
Since calculations given in this section are rather complicated and almost the
same procedures as the former section \ref{tumbling} are taken, we would like
to describe below only the final expressions. The detailed calculations are
given in the Appendix.

The tree action for $SU(5)$ massless gauge theory is already given by
Eq.~(\ref{S0}). We consider a symmetry breaking pattern: 
$SU(5)\rightarrow SU(4)$. Each multiplet in $SU(5)$ is decompsed into a sum of
$SU(4)$ multiplets as
\begin{eqnarray}
  \underline{5}^* &=& \underline{4}^*+\underline{1}, \qquad
  \underline{10} = \underline{6}+\underline{4},\qquad
  \underline{24} = \underline{15}+\underline{4}+\underline{4}^*
  +\underline{1} \label{gauge}
\end{eqnarray}
Then $\underline{4}$, $\underline{4}^*$, and \underline{1} of gauge bosons
included in \underline{24} of Eq.~(\ref{gauge}) become massive after
\underline{10} fermions are condensed. Our final purpose is to show the method
how to obtain masses for these gauges bosons.

We follow the same procedures taken in Sect. \ref{tumbling} except for that
1) instead of functionally integrating over all the gauge fields we integrate
over only {\sl {massless}} $SU(4)$ gauge fields hence there remain
$9=\underline{4}+\underline{4}^*+\underline{1}$ gauge fields, 2) hence the
coefficient of the
scalar component of the Fierz-transformed $J_{10}\times J_{10}$ is different
from that given by Eq.~(\ref{J10J10}), i.e., that coefficient given in
Eq.~(\ref{J10J10}) is only used to which channel the symmetry breaks down (MAC),
and 3) dynamical Higgs is from the beginning decomposed into 
$\underline{1}+\underline{4}$. Leaving all the details to the Appendix
\ref{Leff}, we give only the final expression for the effective action which
gives the Feynman rules among the would-be massive gauge bosons, the
Nambu-Goldstone bosons, and fermions
\begin{eqnarray}
  S_{\rm eff}&=&
  \int {d^4 x} \Biggl\{
  {\frac{1}{4} { \left(\overline
  \Psi_{\underline 6}\right)^{p\,q}} (x)\,i\left[ {{\partial\kern-0.6em /} \,
  - \frac{1}{\sqrt{10}}ig\gamma_5{G\kern-0.7em /}\;(x)} \right]\,
  \left(\Psi_{\underline 6}\right)_{qp}(x)}+
  \left(\overline\Psi_{\underline 4}\right)^p (x)\,i
  {\partial\kern-0.6em /} \left(\Psi_{\underline 4}\right) _p (x)
  \nonumber \\
  &+& \left(\overline\Psi_{\underline 6}\right)^{q\,p} (x)\,
  g{F\kern-0.6em /}{~}^\rho (x)\left( {S^\rho } \right)_q {}^5
  \,L\,\left(\Psi_{\underline 4}\right)_p (x)-
  \left(\overline\Psi_{\underline 4}\right)^p (x)\,
  g{{F\kern-0.6em /}{~}^\rho}{~}^\dagger(x)
  \left( {S^\rho }{~}^\dagger \right)_5 {} ^q\,L\,\left(\Psi_{\underline 6}
  \right)_{p\,q} (x)\Biggr\}
  \nonumber \\
  &+& \int{d^4 x}{d^4 y}\,\Biggl\{ -\left(\overline\Psi_{\underline 6}
  \right)^{p\,q} (x)\,\phi_0(x-y)\,\left(\Psi_{\underline 6}\right)_{qp}(y)
  + 3\sqrt{2} i\,\phi_0(x-y)\times \nonumber \\
  & \times & \Biggl[\frac{1}{6v_1} \varphi^0(X)
  \left(\overline\Psi_{\underline 6}\right)^{p\,q}(x)\,\gamma_5\,
  \left(\Psi_{\underline 6}\right)_{q\,p}(y)
  + \frac{1}{v_2}\left(\overline\Psi_{\underline 6}\right)^{q\,p} (x)\,
  \varphi^\rho(X) \left({S^\rho  } \right)_q {} ^5\,L\,
  \left(\Psi_{\underline 4}\right)_p (y)
  \nonumber \\
  &+& \frac{1}{v_2}\left(\overline\Psi_{\underline 4}\right)^p (x)\,
  {\varphi^\rho}{~}^\dagger(X)
  \left( {S^\rho }{~}^\dagger \right)_5 {} ^q\,R\,\left(\Psi_{\underline 6}
  \right)_{p\,q} (y) \Biggr]-\frac{1}{2}\phi_i^\dagger(x,y)
  D_1^{-1}(x-y)\phi_i(x,y) \Biggr\}+\ldots,
  \label{Seff}
\end{eqnarray}
where $X=(x+y)/2$, $R,\;L=(1\pm \gamma_5)/2$, $\rho=1 \sim 4$, $D_1(x-y)$ is
defined in the Appendix Eq.~(\ref{D1}), and we have
replaced fields $A_\mu^\alpha$ and $\pi^\alpha$, and generators $T^\alpha$ of
indices $\alpha=16\sim 24$ with the followings.
\begin{mathletters}
\begin{eqnarray}
  F_\mu^1(x) &\equiv& \frac{A_\mu^{16}(x)-iA_\mu^{17}(x)}{\sqrt{2}}
  \quad{\rm etc.} \ldots,
  \quad G_\mu(x) \equiv A_\mu^{24}(x),
  \quad {{F\kern-0.6em /}{~}^\rho}{~}^\dagger(x)
  \equiv \gamma^\mu \left(F_\mu^\rho (x)\right)^\dagger, \\
  \varphi^1(X) &\equiv& \frac{\pi^{16}(X)-i\pi^{17}(X)}{2}
  \quad{\rm etc.} \ldots,
  \quad \varphi^0(X) \equiv \sqrt{2}\left(T^{24}\right)_5{}^5 \pi^{24}(X),
  \label{varphi} \\
  \left( {T^{24} } \right)_r{\kern 1pt}^s &=& \frac{1}{2\sqrt{10}}
  \;\delta_r{\kern 1pt} ^s\quad{\rm for~}r,~s=1{\rm ~\sim 4},\qquad
  \left( {T^{24} } \right)_5{\kern 1pt}^5=-\frac{2}{\sqrt{10}}, \\
  S^1 &\equiv& \frac{T^{16}+i T^{17}}{\sqrt{2}}
  \quad{\rm etc.} \ldots.
\end{eqnarray}
\end{mathletters}
Here the particles included in the action Eq.~(\ref{Seff}) are
given by the \underline{4} and \underline{1} massive gauge bosons
$F_\mu^\rho(x)$ and $G(x)$, the corresponding Nambu-Goldstone bosons,
$\varphi^\rho(x)$ and $\varphi^{0}(x)$, and massive {\underline 6} and
massless {\underline 4} dirac fermions $\Psi_{\underline 6}(x)$ and
$\Psi_{\underline 4}(x)$, respectively, defined by
\begin{mathletters}
\begin{eqnarray}
  \left(\Psi_{\underline 6}\right)_{p\,q}(x)&=&\chi_{p\,q}(x)+\frac{1}{2}
  \epsilon_{p\,q\,r\,s}\,U_c\,{\overline{\chi^{~}_{r\,s}}}(x)^t, \\
  \left(\Psi_{\underline 4}\right)_p(x)&=&\chi_{p\,5}(x)+\psi_p(x).
\end{eqnarray}
\end{mathletters}
where a superscript $t$ is a transposed operator on spinor indices.
Assumed is the following decomposition of five-dimensional bilocal field
$\phi_i(x,y)$ into singlet
$\tilde\sigma=v_1+(\sigma+i \varphi^{0})/\sqrt{2}$ and quartet $\varphi^\rho$
($\rho=1\sim 4$) introduced in Eqs.~(\ref{Seff}, \ref{varphi}) with two
generally different massive parameters $v_i$.
\begin{equation}
  \phi_i(x,y) = \frac{\phi_0(x-y)}{v_1}\left[v_1+
  \frac{\sigma (X)+i\varphi^0(X)}{\sqrt{2}}\right]\delta_i{}^5
  +\sqrt{2} i\frac{\phi_0(x-y)}{v_2}\varphi^\rho (X)
  \left(S^\rho \right)_i{}^5+\ldots, \label{phi1_4}
\end{equation}
with $\alpha=16\sim 23$ and $\ldots$ means the rest. Notice this expression of
$\phi_i(x,y)$ does not include all the
condensation information in it since we have neglected contributions from
massive gauge fields though they also belong to $SU(5)$ gauge group. This causes
the softly massive NG bosons, i.e., the gap equation for the {\underline 6}
fermion multiplet does not assure masslessness of the NG bosons $\varphi^0$ and
$\varphi^\rho$ anymore. These become the pseudo-NG bosons whose mass is
suppressed as the order of 1/(gauge boson mass)$^2$. The two parameters $v_1$
and $v_2$ correspond to decay constants of $G_\mu$ and $F_\mu^\rho$,
respectively.

After obtaining the effective action Eq.~(\ref{Seff}), it is an easy task to
extract Feynman rules and calculate decay constans or two point functions
between gauge bosons and NG bosons through fermion one-loop diagrams which
determine gauge boson masses as usual that will be done in the next section.

%%%%%%%%%%%%%%%%%%%%%%%%%%%%%%%%%%%%%%%%%%%%%%%%%%%%%%%%%%%%
%    section mass Expression
%%%%%%%%%%%%%%%%%%%%%%%%%%%%%%%%%%%%%%%%%%%%%%%%%%%%%%%%%%%%
\section{Mass Formulae}
\label{mass}
\begin{center}
%
% Figure 2
%
\begin{figure}
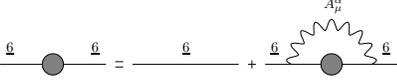

\center\psbox[hscale=0.5,vscale=0.5]{nfig2.eps}
\caption{Gap equation for {\underline 6} fermion.}
\label{Fig2}
\end{figure}
\end{center}

Due to the arguments in the former Sections, the Feynman diagrams of
the simultaneous gap equations for {\underline 6} fermion, {\underline 4}, and
{\underline 1} gauge bosons can be depicted in Figs. \ref{Fig2} and \ref{Fig3}.

\begin{center}
%
% Figure 3
%
\begin{figure}
\center\psbox[hscale=0.5,vscale=0.5]{nfig3.eps}
\caption{Gap equation for {\underline 4} ( {\underline 1} ) gauge bosons.}
\label{Fig3}
\end{figure}
\end{center}

Calculating Fig. \ref{Fig2}, the gap equation for the {\underline 6} fermion is
given by
\begin{equation}
  m_6(x)=-4\int d^4r\; e^{-ip\,r}\phi_0(r)
  =6 \lambda \,\int dz\frac{zm_6(z)}{z+m_6(z)^2}
  \left[\frac{\theta (z-x)}{z}+ \frac{\theta (x-z)}{x}\right]
  ,\label{m6}
\end{equation}
with
\begin{equation}
  \lambda=\pi g^2C_D\lambda_1=\frac{\pi g^2}{2^D\pi^{(D+1)/2}\Gamma
  \left(\frac{D-1}{2}\right)}\lambda_1=\frac{g^2}{8\pi^2}\lambda_1 ,\qquad
  \lambda_1=\frac{3}{N_5}\left(1-\frac{1}{N_4^2}\right)
  =\frac{9}{16^2}, \label{lambda} \\
\end{equation}
where $m_6(x)$ is mass for {\underline 6} fermion, $g$ is a gauge coupling
constant, $x=-p^2$, the space-time dimension is given by $D=4$, $N_4=4$, and
$N_5=5$. Eq.~(\ref{m6}) is a well-known formula up to a constant
and the structure of the solution is well studied. Here we have neglected
contributions from massive gauge bosons which have been a hot subject known as
a gauged Nambu-Jona-Lasinio model\cite{GNJ}. Note that the coefficent of the rhs
of Eq.~(\ref{m6}) is quite different from that obtained in Eq.~(\ref{J10J10})
( $\propto 2C_{10}-C_5$ ) expected from the tumbling.

\begin{center}
%
% Figure 4
%
\begin{figure}
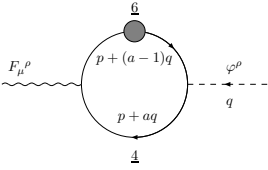

\center\psbox[hscale=0.5,vscale=0.5]{nfig4.eps}
\caption{Fermion one-loop contribution to $F_\mu^\rho$ - $\varphi^\rho$
two point function.}
\label{Fig4}
\end{figure}
\end{center}

\begin{center}
%
% Figure 5
%
\begin{figure}
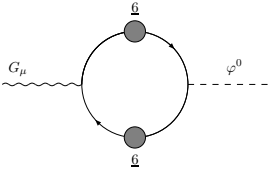

\center\psbox[hscale=0.5,vscale=0.5]{nfig5.eps}
\caption{Fermion one-loop contribution to $G_\mu$ - $\varphi^0$
two point function.}
\label{Fig5}
\end{figure}
\end{center}

After calculating two-point functions between $F^\rho_\mu(x)$ 
({\underline 4}) and $\varphi^\rho(x)$ and/or $G_\mu(x)$ ({\underline 1}) and
$\varphi^{0}(x)$ depicted in Figs.~\ref{Fig4} and \ref{Fig5}, and also using
Fig.~\ref{Fig3}, i.e., Figs.~\ref{Fig4} and \ref{Fig5} devided by $i$ being
equated to $-ig\;v_2\;q_\mu/\sqrt{2}$ and $-2ig\;v_1\;q_\mu/\sqrt{5}$\cite{MS},
respectively, masses for the gauge bosons are given in Euclidean space by
\begin{mathletters}
\label{massFG}
\begin{eqnarray}
  M_F^2 &=& \left(\frac{gv_2}{\sqrt{2}}\right)^2=
  \frac{9 g^2}{2}\int\;\frac{dx}{16\pi^2}\frac{1}{x+m_6^2(x)}\Biggl\{
  3xm_6(x) m'_6(x)-\frac{2x m_6^2(x)\left(1+2 m_6(x) m'_6(x)\right)}
  {x+m_6^2(x)} \nonumber \\
  &-& 2a\;\left[xm_6(x) m'_6(x)-\frac{x m_6^2(x)\left(1+2 m_6(x) m'_6(x)\right)}
  {x+m_6^2(x)} - m_6^2(x) \right]\Biggr\}, \\
  M_G^2 &=& 2\left(\frac{2gv_1}{\sqrt{10}}\right)^2 = 
  \frac{24 g^2}{5} \int\;\frac{x dx}{16\pi^2}
  \frac{m_6(x)\left(m_6(x)-\frac{1}{2} x m'_6(x)\right)}
  {\left(x+m_6^2(x)\right)^2},
\end{eqnarray}
\end{mathletters}
where $a$ is an arbitrary parameter which comes in the fermion loop
(see Fig.~\ref{Fig4}). Values of the vertices on the left hand side of 
Fig.~\ref{Fig3}, i.e., $-ig\;v_2\;q_\mu/\sqrt{2}$ and $-2ig\;v_1
\;q_\mu/\sqrt{5}$, are given by the kinetic terms of gauge bosons and NG bosons
in Eqs.~(\ref{kineticD}) of the next Section, which are invariant under $SU(4)$.

$M_G$ can be determined uniquely {\it a la} Pagels and
Stokar once $m_6(x)$ is solved, while $M_F$ cannot because of one arbitrary
parameter $a$ which determines distribution of internal momentum in the fermion
loop. The physical results should not depend on this kind of arbitrary
parameters. One reason for this is that the right hand side vertices in
Figs.~\ref{Fig4} and \ref{Fig5} depend on the form factor $m_6(-p^2)$ which is
not a point interaction. However $M_G$ is determined uniquely even though it
has the same situation as $M_F$. Another reason, which will be explained
more in the next Section, may be that the Ward-Takahashi (WT) identity holds for
the fermion current of the  {\underline 6}-{\underline 6}-$G_\mu$ vertex in
Fig.~\ref{Fig5}, while it does not hold for the current of
{\underline 6}-{\underline 4}$^*$-$F_\mu^\rho$ vertex in Fig.~\ref{Fig4}.

%%%%%%%%%%%%%%%%%%%%%%%%%%%%%%%%%%%%%%%%%%%%%%%%%%%%%%%%%%%%
%    section Summary and Discussions
%%%%%%%%%%%%%%%%%%%%%%%%%%%%%%%%%%%%%%%%%%%%%%%%%%%%%%%%%%%%
\section{Summary and Discussions}
\label{discuss}
In this paper we have shown that taking an example of $SU(5)$ massless gauge
theory dynamical breakdown of gauge symmetry yields different results with
regards to gauge boson masses when compared with elementary Higgs. {\sl Even
though
it is not proven, we believe that this is not a feature peculiar to $SU(5)$
theory but rather a general phenomenon.} If we have an elementary Higgs, then
the relevant kinetic terms of {\underline 4}/${\underline 4}^*$ and
{\underline 1} would-be massive gauge bosons would be given by
\begin{mathletters}
\label{kineticH}
\begin{eqnarray}
  {\cal L}_\phi &=& \left| D _\mu\;\varphi^\rho-\frac{g}{\sqrt{2}}\;
  \tilde\sigma F_\mu^\rho 
  \right|^2 +\left| D _\mu \tilde\sigma \right|^2, \\ %\quad
  \tilde\sigma &=& v + \frac{\sigma+i \varphi^0}{\sqrt{2}}, \quad %\\
  D _\mu \varphi^\rho = \left( \partial _\mu -\frac{ig}{2\sqrt{10}} G_\mu
  \right) \varphi^\rho, \quad
  D _\mu \tilde\sigma = \left( \partial _\mu +\frac{2ig}{\sqrt{10}} G_\mu
  \right)\tilde\sigma.
\end{eqnarray}
\end{mathletters}
This is the case
when one decomposes $SU(5)$ gauge invariant kinetic term of 5-dimensional
elementary Higgs, $\phi_i$ into ${\underline 4}+{\underline 1}$ and gauge
fields, $A_\mu^A$ into ${\underline {15}}+{\underline 4}+{\underline 4}^*+
{\underline 1} (={\underline {24}})$ under $SU(4)$,
among which {\underline 4}, ${\underline 4}^*$, and {\underline 1}, i.e.,
$F_\mu^\rho$, its complex comjugates, and $G_\mu$ become massive. Their masses
are given by
\begin{equation}
  M_F^2 = \left(\frac{g v}{\sqrt{2}}\right)^2, \quad 
  M_G^2 = 2\left(\frac{2g v}{\sqrt{10}}\right)^2, \quad
  \left(\frac{M_G}{M_F}\right)^2 = \frac{8}{5}. \label{massE}
\end{equation}

Comapared with the expression Eq.~(\ref{kineticH}), we have seen in the former
section that there are two different decay constants or two different mass
deimensionful parameters, $v_1$ and $v_2$. Both of these cannot be VEV's
at the same time since only one scalar $\tilde\sigma$ is available. Contrary to
the case where an elementary Higgs is included from the beginning, we need, in
the case of dynamical theory, only $SU(4)$ invariant kinetic terms. Then we may
consider the following expression for the kinetic terms which give different
gauge boson masses for $F_\mu^\rho$ and $G_\mu$ and agree with what we have
obtained in the former sections.
\begin{equation}
  {\cal L}_\phi = \left| D _\mu\;\varphi^\rho-\frac{g}{\sqrt{2}}
  \frac{v_2}{v_1}\;\tilde\sigma F_\mu^\rho \right|^2
  +\left| D_\mu \tilde\sigma \right|^2, \label{kineticD}
\end{equation}
with $\tilde\sigma=v_1 + (\sigma + i \varphi^0)/\sqrt{2}$. With this
interpretation, what we have calculated in the former Section are two-point
functions, $\left<0\right|{\rm T}\left(F_\mu^\rho(0)\varphi^\rho(x)\right)
\left|0\right>_{\rm amp}$ and $\left<0\right|{\rm T}\left(G_\mu(0)\varphi^0(x)
\right)\left|0\right>_{\rm amp}$.\cite{MS} In this case, gauge boson masses are
given by
\begin{equation}
  M_F^2 = \left(\frac{g v_2}{\sqrt{2}}\right)^2, \quad 
  M_G^2 = 2\left(\frac{2g v_1}{\sqrt{10}}\right)^2, \quad
  \left(\frac{M_G}{M_F}\right)^2 = \frac{8}{5}\left(\frac{v_1}{v_2}
  \right)^2. \label{massD}
\end{equation}

In sammary, if an elementary Higgs is icluded in Eq.~(\ref{S0}),
Eq.~(\ref{kineticH}) is obtained, on the other hand if massless $SU(5)$
gauge theory is adopted, then Eq.~(\ref{kineticD}) is generated dynamically.
Then the ratio of singlet/quartet gauge boson masses is given by $M_G/M_F=
\sqrt{8/5}$, Eq.~(\ref{massE}), if an elementary Higgs is included, while in
our case it is given by $M_G/M_F=\sqrt{8/5}\;v_1/v_2$, Eq.~(\ref{massD}),
in Sect. \ref{mass} which is a complicated functional of the {\underline 6}
fermion mass $m_6(x)$, Eq.~(\ref{m6}) and also include one arbitrary parameter
$a$.

Actual calculations of $v_i$ show that $v_1$ can be determined just like an
ordinary decay constant while $v_2$ cannot. This is because in the case of the
singlet gauge boson $G_\mu$ the Ward-Takahashi (WT) identity holds for the
axial vector current
\begin{equation}
  \left(\bar\Psi_{\underline 6}\right)^{pq} \gamma_\mu \gamma_5
  \left(\Psi_6\right)_{qp},
\end{equation}
which is the Noether current of the chiral symmety with respect to {\underline
6} representation. On the other hand in the case of the quartet gauge boson
$F_\mu^\rho$ the WT identity does not hold for the current
\begin{equation}
  \left(\bar\Psi_{\underline 6}\right)^{qp} \gamma_\mu L
  \left(S^\rho\right)_q{}^5 \left(\Psi_{\underline 4}\right)_p,
\end{equation}
because this is not the Noether current of the chiral symmetry if gauge bosons
are absent. That is, the fermion current between the same representations
allows to have the WT identiy, while the current between the different
representations does not. This assures that $G_\mu$ is given unique mass, while
$F_\mu^\rho$ is not.

These currents can be rewritten in terms of the original left-handed chiral
fields, $\chi_{i\;j}$,
as
\begin{mathletters}
\begin{eqnarray}
  \overline{\chi_{p\;q}}\;\gamma_\mu\left(T^{24}\right)_q{}^{q'}
  \chi_{q'\;p}, \\
  \overline{\chi_{p\;q}}\;\gamma_\mu\left(S^\rho\right)_q{}^5 \chi_{p\;5},
\end{eqnarray}
\end{mathletters}
respectively, where $\left(T^{24}\right)_q{}^{q'}=\delta_q{}^{q'}/2\sqrt{10}$
for $q,~q'=1\sim 4$.
That is, these are part of full $SU(5)$ ten-dimensional fermion current,
\begin{equation}
  \overline{\chi_{i\;j}}\;\gamma_\mu\left(T^{\alpha}\right)_j{}^{j'}
  \chi_{j'\;i}.
\end{equation}
With this current, it seems that one could have the WT identity with one
decay constant. However this does not hold as we have seen in Sect. 
\ref{tumbling} since two kinds of the fermion loop structures appear and they
give different decay constants after gauge symmetry breaks down from $SU(5)$ to
$SU(4)$. This is part of the reason why we have an arbitrary constant $a$ for
a decay constant of the quartet gauge bosons.

In this paper, we have pointed out existence of one possibility how to caculate
gauge boson masses at the same time fermions acquire mass. However, we have
not yet succeeded in finding any principle either to fix the parameter $a$ or
to construct the fully consistent method to calculate
gauge boson masses as well as fermion masses. This might be
in a sence in the same situation as a gauge parameter $\alpha_0$, on which
masses generally depend as you can see easily from the fermion gap equation
although physical qunatities should not depend on it. These may be a future
problem to be solved in many models including super symmetric ones.

\acknowledgements
M. S. acknowledges Prof. T. Morii for his continuous encouragements during the
course of this study. T. M. and M. S. would also like to thank the theory
group of KEK at Tanashi which now moves to Tuskuba where part of this work has
been done. T. M. would also like to acknowledge Prof. W. Haymaker for his
hospitality at Louisiana State University where part of this work has been
done.

%%%%%%%%%%%%%%%%%%%%%%%%%%%%%%%%%%%%%%%%%%%%%%%%%%%%%%%%%%%%
%    section Appendix
%%%%%%%%%%%%%%%%%%%%%%%%%%%%%%%%%%%%%%%%%%%%%%%%%%%%%%%%%%%%
\appendix
%%%%%%%%%%%%%%%%%%%%%%%%%%%%%%%%%%%%%%%%%%%%%%%%%%%%%%%%%%%%
%    section Effective Lagrangian
%%%%%%%%%%%%%%%%%%%%%%%%%%%%%%%%%%%%%%%%%%%%%%%%%%%%%%%%%%%%
\section{Effective Lagrangian}
\label{Leff}
In this Appendix, we will derive the effective Lagrangian, Eq.~(\ref{Seff}).
Given the starting action Eq.~(\ref{S0}), we first partially fix the gauge
for gauge fields $A^A_\mu$ for $A=a=1\sim 15$ belonging to the $SU(4)$
sub-group, functionally integrate the partition function over these gauge
fields, and then obtain the effective action as
\begin{eqnarray}
  S_1&=&\int d^4x \Biggl\{i\,\left(\overline\psi\right)^i\left[
  {\partial\kern-0.6em /}\;\delta{}_i{}^j-ig{A^\alpha\kern-1.0em /}
  {\kern+0.7em}\left(T^\alpha\right)_i{}^j\right]\psi_j
  +i\;\frac{1}{2}\left(\overline \chi\right)^{i\;j}\left[
  {\partial\kern-0.6em /}{\kern+0.4em}\delta_j{}^k-2ig{A^\alpha\kern-1.0em /}
  {\kern+0.7em}\left(T^\alpha\right)_j{}^k\right]\chi_{k\;i}
  \nonumber \\
  &-&\frac{1}{4}\left(\partial_\mu\,A_\nu^\alpha
  -\partial_\nu\,A_\mu^\alpha\right)^2\Biggr\}
  +\frac{1}{2}g^2\int d^4x\;d^4y\;\Big[{J_5}^a_\mu(x)
  \delta_{a\;b}D^{\mu\;\nu}(x-y){J_5}^b_\nu(y)
  \nonumber \\
  &+&{J_{10}}^a_\mu(x)\delta_{a\;b}D^{\mu\;\nu}(x-y){J_{10}}^b_\nu(y)
  +2{J_5}^a_\mu(x)\delta_{a\;b}D^{\mu\;\nu}(x-y){J_{10}}^b_\nu(y)\Big]
  +\ldots, \label{App:S1}
\end{eqnarray}
where
\begin{mathletters}
\begin{eqnarray}
  {J_5}^a_\mu (x)&=&{\overline\psi}{~}^p(x)\gamma_\mu \left(T^a\right)_p{}^q
  \;\psi_q(x), \\
  {J_{10}}^a_\mu (x)&=&{\left(\overline\chi\right)}{~}^{i\,p}(x)\;\gamma_\mu 
  \left(T^a\right)_p{}^q\;\delta{}_i{}^j\,\chi{}_{q\,j}(x), \\
  D_{\mu\nu}(x)&=&\int\frac{d^4p}{(2\pi)^4}\frac{1}{p^2}
  \left[g_{\mu\nu} -(1-\alpha_0)\frac{p_\mu p_\nu}{p^2}\right]\;e^{-ipx},
\end{eqnarray}
\end{mathletters}
with a superscript $\alpha$ running from 16 through 24, super/subscipts $a$
and $b$ from 1 through 15, $p$ and $q$ from 1 through 4 since $T^a$ are $SU(4)$
generators, $i$, $j$, and $k$ from 1 through 5, and $\alpha_0$ being a
gauge parameter for $SU(4)$ in the last equation. $A_\mu^\alpha$ are the
would-be massive gauge fields. Expessed in Eq.~(\ref{App:S1}) are only the
contributions from ladder approximation and $\ldots$ are the remains. Necessary
four-fermi terms are obtained from Fierz-transforming only ${J_{10}}_\mu^a(x)
{J_{10}}_\nu^a(x)$ term since ${J_5}_\mu^a(x){J_5}_\nu^a(x)$ does not include
scalar modes and ${J_5}_\mu^a(x){J_{10}}_\nu^a(x)$ does not develop a VEV
compared with $J_{10}J_{10}$. The last statement can be easily checked when one
compares the gap equations for the fermions {\underline 4} and {\underline 6}
and sees that the coupling for {\underline 6} is larger than
{\underline 4}. This is exactly what the MAC gives us in the $SU(5)$ case.
Hence here we write down only the Fierz-transformed ${J_{10}}_\mu^a(x)
{J_{10}}_\nu^a(y)$.
%
%\begin{mathletters}
\begin{eqnarray}
  &{~}&{J_{10}}^a_\mu (x){J_{10}}^a_\nu (y) \nonumber \\
  &=&-\frac{1}{8}\;{\rm Tr}\;\left[{\left(\overline\chi\right)}{~}^{i\;p}(x)\gamma_\mu 
  \left(T^a\right)_p{}^q\;\delta{}_i{}^j\,\chi_{q\;j}(x)\right]
  \;{\rm Tr}\;\left[\left({\tilde\chi{}^{k\ell r}(y)}^T\;
  U_c\right)\;\gamma_\nu\left(T^a\right)_r{}^s\;\delta{}_\ell{}^m\,
  \left(U_c^{-1}{\overline
  {\tilde\chi}}{}_{smk}(y)^T\right)\right] \nonumber \\
  &=& \frac{1}{32N_5}\left(1-\frac{1}{\left(N_4\right)^2}\right)\eta_{\mu\nu}\;
  \Biggl\{
  {\rm Tr}\;\left[\left({{\tilde\chi}{~}^{i\,j\,p}}(y)^T\;U_c\right)
  \chi_{p\,j}(x)\right]\;
  {\rm Tr}\;\left[{\left(\overline\chi\right)}{~}^{k\,q}(x)\;\left(U_c^{-1}\;
  {\overline{\tilde\chi}}{}_{q\,k\,i}(y)^T\right)\right]+\ldots, \label{J10J102}
\end{eqnarray}
%\end{mathletters}
%
where Tr means to take a trace over only gamma matrices and a superscript $t$
is a trasnposed operator only on spinor indices, and defined are
\begin{equation}
  \tilde\chi{}^{ijk}(x)=\epsilon^{ijk\ell m}\chi_{\ell m}(x), \qquad
  U_c=i\gamma^0\gamma^2.
\end{equation}
Here use has been made of the Fierz-transformation:
\begin{mathletters}
\begin{eqnarray}
  \left(\gamma_\mu\right)_{\alpha\;\beta}\left(\gamma_\nu\right)_
  {\gamma\;\delta}&=&
  \frac{1}{4}\Biggl\{\eta_{\mu\;\nu}\left[\delta_{\alpha\;\delta}
  \delta_{\gamma\;\beta}-\left(\gamma_5\right)_{\alpha\;\delta}
  \left(\gamma_5\right)_{\gamma\;\beta}\right]
  \nonumber \\
  &-&\eta^{\rho\;\sigma}\left[\left(\sigma_{\rho\;\mu}\right)_{\alpha\;\delta}
  \left(\sigma_{\sigma\;\nu}\right)_{\gamma\;\beta}-\left(\gamma_5
  \sigma_{\rho\;\mu}\right)_{\alpha\;\delta}\;
  \left(\gamma_5\sigma_{\sigma\;\nu}\right)_{\gamma\;\beta}\right]\Biggr\},\\
  \delta_i{}^j\;\delta_k{}^\ell&=&
  \frac{1}{N_5}\,\delta_i{}^\ell\,\delta_k{}^j+2\left(T^A\right)_i{}^\ell\,
  \left(T^A\right)_k{}^j, \\
  \left(T^a\right)_p{}^q\,\left(T^a\right)_r{}^s&=& \frac{1}{2}
  \left(1-\frac{1}{\left(N_4\right)^2}\right)\delta_p{}^s\;\delta_r{}^q-
  \frac{1}{N_4}\,\left(T^a\right)_p{}^s\,\left(T^a\right)_r{}^q,
\end{eqnarray}
\end{mathletters}
where several kinds of indices are explained before and $N_m$ is a dimension of
$SU(m)$, i.e., $N_m=m$. By taking into account that $\gamma_5\,
\Psi_L=-\Psi_L$, we obtain Eq.~(\ref{J10J102}). Now introducing the following
quadratic auxiliary field term to Eq.~(\ref{App:S1}),
\begin{eqnarray}
  S_{\rm AF}&=&-\frac{1}{2}
  \int\,\int\,d^4x\,d^4y\,\Biggl\{\left[
  {\phi_i}{}^\dagger(x,y)-
  \left({{\tilde\chi}{~}^{i\,j\,p}}(y)^T\;U_c\right)
  \chi_{p\,j}(x)\,D_1(x-y)\right]
  \nonumber \\
  &\times& D_1^{-1}(x-y)\;\left[
  \phi_i(x,y)-
  D_1(x-y)\,{\left(\overline\chi\right)}{~}^{k\,q}(x)\;
  \left(U_c^{-1}\;{\overline{\tilde\chi}}{}_{q\,k\,i}(y)^T\right)\right] 
  +\ldots, \label{App:Saf}
\end{eqnarray}
we obtain
\begin{eqnarray}
  S_3&=&S_2+S_{\rm AF}
  \nonumber \\
  &=&\int d^4x \Biggl\{i\;\left(\overline\psi\right)^i\left[
  {\partial\kern-0.6em /}\;\delta_i{}^j-ig{A^\alpha\kern-1.0em /}
  {\kern+0.7em}\left(T^\alpha\right)_i{}^j\right]\psi_j
  +i\;\frac{1}{2}\left(\overline \chi\right)^{i\;j}\left[
  {\partial\kern-0.6em /}{\kern+0.4em}\delta_j{}^k-
  2ig{A^\alpha\kern-1.0em /}
  {\kern+0.7em}\left(T^\alpha\right)_j{}^k\right]\chi_{k\;i}
  \nonumber \\
  &\quad&-\frac{1}{4}\left(\partial_\mu\,A_\nu^\alpha
  -\partial_\nu\,A_\mu^\alpha\right)^2\Biggr\}
  \nonumber \\
  &+&\frac{1}{2}\int\,d^4x\,d^4y\,\left[\left({{\tilde\chi}{~}^{i\,p\,j}}
  (y)^T\;U_c\right) \phi_i(x,y)\;\delta{}_j{}^k\;
  \left(T^\alpha\right){}_j{~}^k \chi_{k\,p}(x)+{\rm H.C.}\right],
  \nonumber \\
  &-&\frac{1}{2}\int\,d^4x\,d^4y\,
  {\phi_i}{}^\dagger(x,y)\;D_1^{-1}(x-y)\;\phi_i(x,y)+\ldots, \label{App:S3}
\end{eqnarray}
where only relevant terms are written ( $\alpha=16\sim24$ ), and
\begin{mathletters}
\label{D1}
\begin{eqnarray}
  D_1(x-y)&=&\lambda'_1\,D(x-y), \qquad
  \lambda'_1=\frac{1}{16N_5}\left(1- \frac{1}{\left(N_4\right)^2}\right) \\
  D(x-y)&=&\frac{g^2}{4}\,g^{\mu\,\nu}D_{\mu\,\nu}(x-y)=
  \frac{3+\alpha_0}{4}\,g^2\int\frac{d^4p}{\left(2\pi\right)^4}
  \frac{e^{-ip(x-y)}}{p^2}=-\frac{(3+\alpha_0)g^2}{16\pi^2}
  \frac{1}{\left(x-y\right)^2},
\end{eqnarray}
\end{mathletters}
where $\alpha_0$ is a gauge parameter and function $D^{-1}_1(x-y)$ is equal
to $1/D_1(x-y)$.

Now we introduce new fields which make the effective Lagrangian
Eq.~(\ref{App:S3}) much simpler form and assume the forms of auxiliary
fields as
\begin{mathletters}
\begin{eqnarray}
  \left(\Psi_{\underline 6}\right)_{p\,q}(x)&=&\chi_{p\,q}(x)+\frac{1}{2}
  \epsilon_{p\,q\,r\,s}\,U_c\,{\overline\chi}{~}^{r\,s}(x)^t, \\
  \left(\Psi_{\underline 4}\right)_p(x)&=&\chi_{p\,5}(x)+\psi_p(x), \\
  \phi_5(x,y)&=& \frac{\phi_0(x-y)}{v_1}\left[v_1+
  \frac{\sigma\left(\frac{x+y}{2}\right)+
  \sqrt{2} i \left(T^{24}\right)_5{}^5 \pi^{24}\left(\frac{x+y}{2}\right)}
  {\sqrt{2}} \right] \\
  \phi_p(x,y)&=& i \frac{\phi_0(x-y)}{v_2}\,
  \left(T^\alpha\right)_p{}^5 \pi^\alpha\left(\frac{x+y}{2}\right),
\end{eqnarray}
\end{mathletters}
where $p$ and $q$ run from 1 through 4 and $\chi_{p\,q}(x)$ and
$\chi_{p\,5}(x)$ are left handed, while ${\overline\chi}{~}^{r\,s}(x)^t$
and $\psi_p(x)$ are right handed fields as stated in the first part of
this section, and the superscript $t$ stands for the transposed operator for
the spinor indices. Fields with arguments $(x+y)/2$ are dynimical fields,
while those with $x-y$ are classical ones. $\phi_0(x-y)$ is a vacuum
expectation value (VEV) of $\phi_5(x,y)$ and $v$ is a VEV of a scalar
field defined by $v_1\,\phi_5(x,y)/\phi_0(x-y)$. The field $\sigma(X)$ is
a would-be dynamical massive Higgs scalar and the field $\varphi^\alpha(X)$
is a would-be dynamical Nambu-Goldstone boson. Note also that
\begin{eqnarray*}
  {\left(\Psi_{\underline 6}{}^T(x)\right)_{p\,q}}&=&\chi_{q\,p}(x){}^t+
  \frac{1}{2}\,\epsilon_{q\,p\,r\,s}\,U_c{}^t\,{\overline\chi}{~}^{r\,s}(x)
  =-\chi_{p\,q}(x){}^t+\frac{1}{2}\,\epsilon_{p\,q\,r\,s}\,U_c\,
  {\overline\chi}{~}^{r\,s}(x),
\end{eqnarray*}
where $T$ stands for the transposed operator for all indices, while $t$ only for
spinor indices. With these fields the effective Lagrangian $S_{\rm eff}$
Eq.~(\ref{App:S3}) can be given by Eq.~(\ref{Seff}).
%

%%%%%%%%%%%%%%%%%%%%%%%%%%%%%%%%%%%%%%%%%%%%%%%%%%%%%%%%%%%%
%    section Derivation of Masslessness of Nambu-Goldstone bosons
%%%%%%%%%%%%%%%%%%%%%%%%%%%%%%%%%%%%%%%%%%%%%%%%%%%%%%%%%%%%
%\section{Derivation of Masslessness of Nambu-Goldstone bosons}
%\label{massless}
%To prove masslessness of the NG bosons appearing in Eq.~(\ref{App:S3}) or
%Eq.~(\ref{Seff}), we assume the following decomposition of bilocal fields
%first introduced as auxiliary fields.\cite{MS}
%%
%\begin{mathletters}
%\begin{eqnarray}
%  \phi_5(x,y)&=& \phi_0(x-y)+\frac{\sigma_0(x-y)}{v_1}
%  \sigma\left(\frac{x+y}{2}\right) \\
%
%  \phi_p(x,y)&=& i \frac{\pi_0(x-y)}{v_2}\,
%  \left(T^\alpha\right)_p{}^5 \pi^\alpha\left(\frac{x+y}{2}\right),
%\end{eqnarray}
%\end{mathletters}
%
%where only $SU(4)$ invariance is assumed, i.e., $\phi_5$ is singlet and
%$\phi_p$ quartet, so that values of two parameters, $v_1$ and $v_2$, with mass
%dimensions could be in general different.

%The gap equation for $\phi_0(x-y)$ is given by Eq.(\ref{m6}) in a momentum
%space, which gives mass to {\underline 6} fermions. It can be rewritten as
%
%\begin{equation}
%  \partial_q^2\;m_6(q)+\frac{4i\lambda_6(q)}{q^2-m_6(q)}=0
%\end{equation}
%%
%in a Minkowski space since
%
%\begin{eqnarray}
%  m_6(q) &=& 4i\lambda\int\;\frac{d^4p}{(2\pi)^2}
%  \frac{1}{(p-q)^2}\frac{m_6(p)}{p^2-m_6(p)^2}, \\
%
%  \left(\partial_q^2\right)^{-1} &=& -\int\;\frac{d^4p}{(2\pi)^2}
%  \frac{1}{(p-q)^2}.
%\end{eqnarray}
%

%%%%%%%%%%%%%%%%%%%%%%%%%%%%%%%%%%%%%%%%%%%%%%%%%%%%%%%%%%%%
%    References
%%%%%%%%%%%%%%%%%%%%%%%%%%%%%%%%%%%%%%%%%%%%%%%%%%%%%%%%%%%%

\end{document}